\documentclass[%
 amsmath,amssymb,
 aps,prl,twocolumn
]{revtex4-2}

\usepackage{graphicx}
\usepackage{dcolumn}
\usepackage{bm}
\usepackage{color}

\newcommand{\wait}{\mathrm{wait}}
\newcommand{\insp}{\mathrm{insp}}
\newcommand{\pick}{\mathrm{pick}}

\newcommand{\jump}{\mathrm{jump}}
\newcommand{\vibr}{\mathrm{vibr}}

\begin{document}

\title{Inspection paradox and jump detection in glassy systems}

\author{Simone Pigolotti$^1$}
\email{simone.pigolotti@oist.jp}
\author{Sándalo Roldán-Vargas$^2$}%
\email{sandalo@ugr.es}
\affiliation{$^1$Biological Complexity Unit, Okinawa Institute of Science and Technology Graduate University, Onna, Okinawa 904-0495, Japan}
\affiliation{%
 $^2$Department of Applied Physics, Faculty of Sciences, University of Granada, 18071 Granada, Spain
}

\date{\today}

\begin{abstract}
Dynamics in glassy systems near the phase transition is characterized by particle jumps. Approaches to describe these dynamics are based on models in which the time and length scales defining the jumps are parameters to be determined. We instead propose a model-independent method to detect these jumps. Our method uses the theory of the inspection paradox to analyze particle trajectories and reveals the time and length scales defining a jump without free parameters. Given its simplicity and generality, our method can be applied to resolve hopping motion in a broad class of systems, including experimental ones. 
\end{abstract}

\maketitle

Microscopic transport in amorphous materials evolves from liquid-like to a hopping dynamics when approaching a phase transition to a disordered solid state \cite{berthier-biroli-review}. This hopping motion, characterized by particle jumps, has been observed in a variety of systems regulated by different control parameters. Examples include: molecular liquids controlled by temperature \cite{Berthier_silica, pinaki_non_gaussian}; colloidal systems ruled by packing fraction \cite{Weeks_colloids}; granular materials controlled by shear stress \cite{Walter_granular}; complex biological media regulated by pH \cite{MIT_lambda13}; and non-equilibrium active systems, where transport is driven by chemical energy of metabolic origin \cite{cell_migration-pnas,cytoplasm-cell}.

Glassy systems provide a paradigmatic example of hopping dynamics \cite{berthier-biroli-review,debenedetti_Stillinger}. Hopping in these systems has been supported by different approaches. Experiments and simulations in which particles are  tracked in space reveal sudden particle displacements that become more prominent as the phase transition is approached \cite{Weeks-Weitz,Ciamarra-jumps,soft_matter_tetrahedral,Ediger-Heterogeneities,berthier-biroli-review}. In the same regime, the Stokes-Einstein fluctuation-dissipation relation, that holds for a liquid and connects viscosity, temperature, and diffusion coefficient, breaks down \cite{Hodgdon-Stillinger, Stokes-Einstein-234D}. Finally, theoretical approaches describe the dynamics in the vicinity of the liquid- to solid-transition as a phenomenon dominated by thermally activated barriers that particles overcome by hopping events \cite{adam_gibbs,RFOT-original,berthier-biroli-review}. 

All of these studies are, directly or indirectly, based on the idea of jump, i.e an event where a particle shows a ``rapid'' and ``large'' displacement. Thus, a precise definition of jump requires two natural scales to be identified: a time scale to understand what we mean by rapid, and a spatial scale to resolve what we mean by large. Up to now, these two scales have been either imposed by ad hoc considerations or by model-dependent assumptions.  For example, the threshold length can be determined based on the idea of cage dynamics (for instance the Debye-Waller factor), while the time scale is usually associated with that of a few particle collisions \cite{Vollmayr-Lee-jumps,Helfferich-jumps,Pastore-jumps,Ciamarra-jumps,Vittoria-jumps}. 

Seminal works \cite{pre-garrahan,translational-decoupling-2,pinaki_non_gaussian} have described the jump dynamics by means of a continuous-time random-walk (CTRW). These models are characterized by two timescales: the one before observing the first jump, and the one between all the subsequent jumps. It was found that jumps increasingly impact the distribution of displacements as the glass transition is approached \cite{pinaki_non_gaussian}. In parallel, the characteristic time of the first jump becomes much longer than that of the subsequent ones \cite{pre-garrahan,translational-decoupling-2,jcp-garrahan,pinaki_non_gaussian}. It was argued that the two timescales differ because, in reality, jumps are not Poissonian events \cite{jcp-garrahan,Ciamarra-jumps}. This argument is based on a statistical phenomenon called the inspection paradox \cite{feller1991introduction}. In a nutshell, the inspection paradox states that, in a stationary time series including intermittent events, the average waiting time from a randomly chosen instant to the next event can be larger than the average time between consecutive events. The inspection paradox is expected to have broader implications for glassy dynamics \cite{pal2022inspection}, that have been explored in a limited way so far.

In this Letter, we apply the theory of the inspection paradox to study jumps in glassy dynamics in a model-independent way. Our main result is a jump-detection method which naturally identifies the spatial and temporal scales defining a jump. We shall see that these physical scales correspond to a ``resonance'' in the relative fluctuations of time intervals between particle jumps. 

We start by briefly introducing the inspection paradox \cite{feller1991introduction,jcp-garrahan,Ciamarra-jumps,pal2022inspection}. To make things concrete, we directly consider as example the statistics of jumps in a particle trajectory. Although jumps have a finite temporal duration, on a long timescale they can be treated as pointlike in time. We therefore assign to each jump $i$ the time $t_i$ at which it has occurred. We assume that the sequence of times $t_i$ is infinite, stationary, and ergodic.  Our main object of interest is the distribution $P_\wait(t)$ of waiting times between two consecutive jumps. We assume that this distribution has finite first and second moments.

\begin{figure}[htb]
\includegraphics[width=8cm]{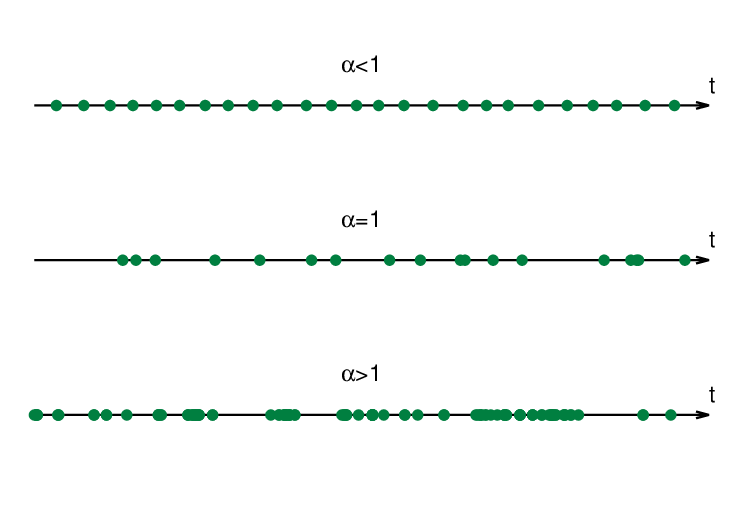}
\caption{Artificial examples of stationary time series of jumps (green circles) with ratio $\alpha=\bar{t}_\insp/\bar{t}_\wait$ smaller, equal, and larger than one.\label{fig:inspect}}
\end{figure}

We now ``inspect'' the time series by choosing a random time with uniform probability. We call $P_\insp(t)$ the distribution of time intervals from the chosen random time to the next jump in the series. Our first goal is to express the distribution $P_\insp(t)$ of the inspection time in terms of the distribution $P_\wait(t)$ of times between consecutive jumps in the series. To this aim, it is convenient to first introduce a third distribution, $P_\pick(t)$, representing the probability that the randomly chosen initial time falls in an interval of duration $t$ between consecutive jumps. Such distribution should be proportional to the density of intervals of duration $t$, which is proportional to $P_\wait(t)$, and the duration $t$ of the interval. This is due to the fact that it is proportionally more likely to pick a long time interval than a short one. This argument leads to conclude that $P_\pick(t)=t P_\wait(t)/\bar{t}_\wait$, where $\bar{t}_\wait\equiv \int_0^\infty dt\, t\, P_\wait(t)$ is the average time between jumps. We now use that the inspection time is uniformly distributed inside an interval of duration $t_\pick$ since any time within this interval is equally likely: $P_\insp(t|t_\pick)=1/t_\pick$. We then have
\begin{equation}
P_\insp(t)=\int_t^\infty dt' P_\insp(t|t_\pick=t') P_\pick(t')\, ,
\end{equation}
and therefore
\begin{equation}\label{eq:feller}
P_\insp(t)=\frac{1}{\bar{t}_\wait}\int_t^\infty dt'\, P_\wait(t').
\end{equation}
Eq.~\eqref{eq:feller} is sometimes referred to as Feller's equation \cite{Ciamarra-jumps}. Computing the average inspection time from Eq.~\eqref{eq:feller} leads to the  relation
\begin{equation}\label{eq:inspection}
\alpha\equiv\frac{\bar{t}_\insp}{\bar{t}_\wait}=\frac{1}{2}\left(1+\frac{\sigma^2_{t_\wait}}{\bar{t}_\wait^{\,2}}\right)
\end{equation} 
where $\sigma^2_{t_\wait}$ is the variance of the distribution $P_\wait(t)$. Equation~\eqref{eq:inspection} shows that the average inspection time $\bar{t}_\insp$ can be smaller, equal, or larger than the average waiting time between jumps $\bar{t}_\wait$ depending on whether $\alpha$ is smaller, equal, or larger than one, respectively, see Fig.~\ref{fig:inspect}. In particular, the marginal case $\alpha=1$ holds when jumps are uniform Poissonian events, so that the distribution $P_\wait(t)$ is exponential. We remark that, although Eq.~\eqref{eq:feller} is usually derived by assuming that the time intervals are independent random variables \cite{feller1991introduction,Ciamarra-jumps,pal2022inspection}, this assumption is not necessary, as we have just shown.

Our aim is to apply this theory to a canonical glassy system. To this aim, we study by molecular dynamics simulations the equilibrium dynamics of the three-dimensional Kob-Andersen binary mixture~\cite{kob-andersen_original} in the microcanonical ensemble. The parameters defining the system interaction are the same as in Ref.~\cite{kob-andersen_original}. We use their same reduced units for length, time, temperature, and energy, with the Boltzmann constant set to one. The temperature $T$ is settled during a previous equilibration process in the canonical ensemble by an Andersen thermostat. We cover a wide range of temperatures $T\in [0.435, 0.9]$: from the liquid state, slightly above the system onset temperature~\cite{Coslovich-Pastore-onset}, down to the estimated mode coupling temperature $T_c = 0.435$~\cite{walter_andersen_PRE}. The numbers of $A$ and $B$ particles in our binary mixture are $N_A = 6400$ and $N_B = 1600$, respectively. The total number density is $\rho = (N_A + N_B)/L^3 = 1.2$, where $L = 18.8$ is the length of the side of the cubic simulation box~\cite{kob-andersen_original}. We used a velocity Verlet algorithm with a time step depending on the temperature as in  Ref~\cite{PRX-BnG}, and ran 5 independent simulations for each value of $T$. Our runs extend from $10^7$ time steps ($10^5$ time units) at high $T$ to $8\cdot10^7$ time steps ($1.6\cdot10^6$ time units) at low $T$. We have verified that these run durations are sufficient to prevent significant finite-time effects. The results presented here correspond to particles of the species $A$.

\begin{figure}[htb]
\includegraphics[width=9cm]{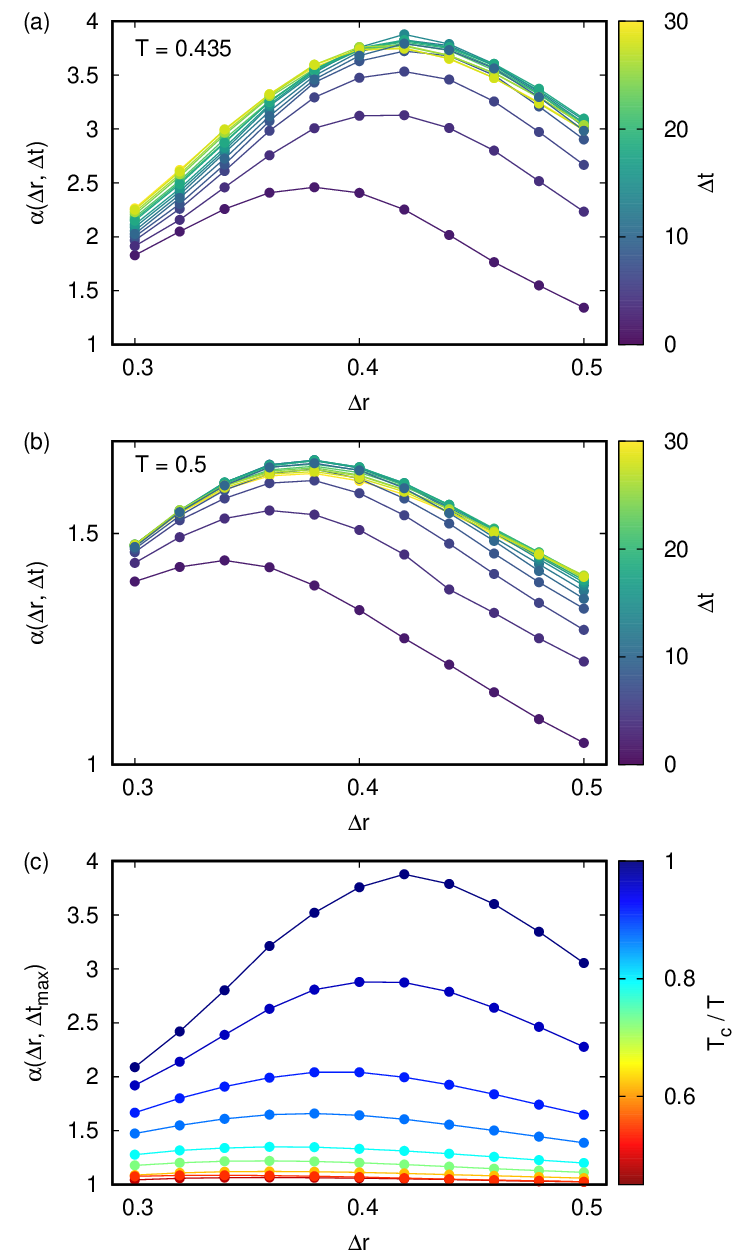}
\caption{Time ratio $\alpha$ as a function of $\Delta r$ and $\Delta t$ at (a) the mode coupling temperature $T_c=0.435$, and (b) an intermediate temperature $T=0.5$. c) $\alpha$ as a function of $\Delta r$  for $\Delta t=\Delta t_{\max}$. The temperatures we investigated are $T \in \{0.435, 0.45, 0.475, 0.5, 0.55, 0.6, 0.7, 0.8, 0.9\}$. In all panels, lines joining the dots are to guide the eye. For each temperature, we covered a range of $\Delta t$ from approximately the system ballistic time \cite{Vittoria-jumps} to a time on the order of several particle collisions. The resolution $\Delta r$ ranges from the estimated length associated to the system vibrational motion \cite{Kob-Andersen-cage} to a distance significantly larger than the typical intracage displacement.\label{fig:alpha-Dr}}
\end{figure}

We define as a jump an event in which a particle travels a net distance greater than $\Delta r$ within a time window $\Delta t$. This simple, model-independent method represents the natural procedure by which an experimental observer would monitor discrete particle trajectories in real space, like in a classical Perrin's experiment \cite{jean-perrin}. It is formulated in terms of two parameters, $\Delta r$ and $\Delta t$, that need to be specified. To analyze the role of these two parameters, we compute the ratio $\alpha$ of characteristic times defined in Eq.~\eqref{eq:inspection} at varying $\Delta r$ and $\Delta t$ for each temperature, see Fig.~\ref{fig:alpha-Dr}a and \ref{fig:alpha-Dr}b. The ratio $\alpha$ presents a maximum as a function of $\Delta r$ and $\Delta t$. This ``resonance'' emerges at all temperatures investigated and becomes more pronounced closer to $T_c$, see Fig.~\ref{fig:alpha-Dr}c.  

The temporal scale $\Delta t_{\max}$  and the spatial scale $\Delta r_{\max}$ corresponding to the resonance naturally emerge from our approach in a model-independent way. Our working hypothesis is that our detection algorithm best captures the heterogeneity of the jump time series for this choice of parameters, meaning that $\Delta t_{\max}$ and $\Delta r_{\max}$ are the scales that best identify the physical jump dynamics.

\begin{figure}[htb]
\includegraphics[width=9cm]{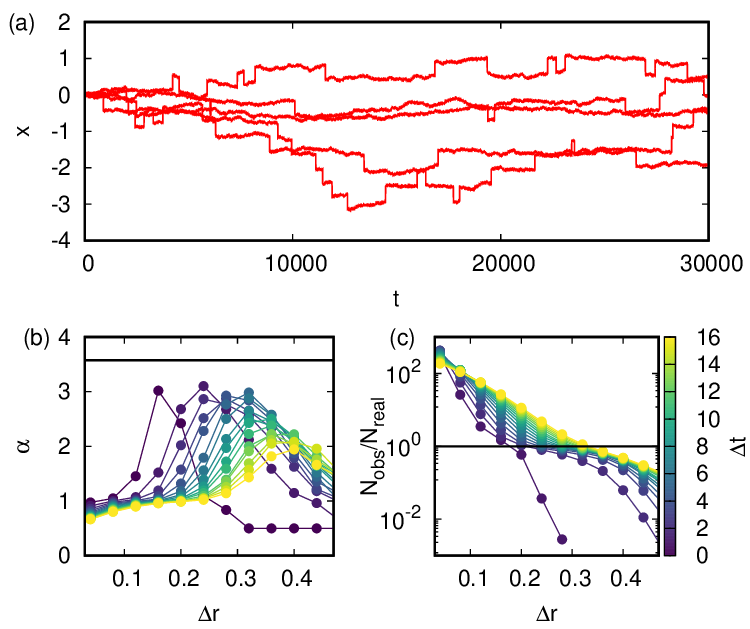}
\caption{Synthetic model. (a) Trajectories of the model. For the interjump time distribution, we choose $t_1=2000$, $t_2=25000$, and $p=0.9$ to mimic the distribution of the Kob-Andersen at $T_c=0.435$ (see \cite{SI}). Other parameters are $\delta r_\jump=0.4$ ($\approxeq \Delta r_{\max}$, Fig.~\ref{fig:alpha-Dr}a), $\delta t_\jump=2$ (time to displace $\delta r_\jump$ ballistically in the Kob-Andersen), $\delta t_\vibr=1$ (ballistic time in the Kob-Andersen \cite{PRX-BnG}), and $\delta r_\vibr=0.017$ (see \cite{SI}). (b) $\alpha$ as a function of $\Delta r$ for different $\Delta t$. The value of the maximum is $\alpha_{\max}\approx 3.1$, compared with the theoretical value $\alpha=3.575$ (horizontal black line). (c) Ratio of observed and real jumps at varying $\Delta r$ and $\Delta t$. In panel (b) and (c), data are obtained by analyzing a single trajectory of the model with $N_{\mathrm{real}}=5000$ jumps.
\label{fig:synthetic}}
\end{figure}

To test this idea and assess how accurately this criterion recovers the actual jumps, we resort to a simple synthetic model, see Fig.~\ref{fig:synthetic}a. In this model, a particle moves in one dimension according to the dynamics 
\begin{equation}
\frac{dx}{dt}=v_\vibr+v_\jump\, .
\end{equation}
The velocity $v_\vibr$ effectively models vibrational motion inside the cage, while the velocity $v_\jump$ models jumps. We call $\delta r_\vibr$ and $\delta t_\vibr$ the characteristic length and time of vibrational motion, respectively. At regular intervals $\delta t_\vibr$, the velocity $|v_\vibr|$ is randomly drawn from an exponential distribution with mean $\delta r_\vibr/\delta t_\vibr$. At the same time, the sign of $v_\vibr$ is drawn at random with uniform probability. We draw the time intervals between jumps from a double-exponential distribution, $P_\wait(t)=p\exp(-t/t_1)+(1-p)\exp(-t/t_2)$, see Fig.~\ref{fig:synthetic}.  When a jump occurs, we set $v_\jump=\pm \delta r_\jump/\delta t_\jump$, where $\delta r_\jump$ and $\delta t_\jump$ are the distance and time associated with a jump, respectively, and we choose the positive/negative sign with equal probability. Each jump lasts for a duration $\delta t_\jump$, with $v_\jump=0$ outside of these time intervals.  

We empirically analyzed a long trajectory of the model, attempting to identify jumps as we did for the Kob-Andersen system. We find that the model presents a maximum in $\alpha$ for values of $\Delta r$ and $\Delta t$ comparable with those of our simulations. The value of $\alpha_{\max}$ is only $14\%$ lower than the actual value of $\alpha$ associated with the interjump time distribution in the model, see Fig.~\ref{fig:synthetic}b. For this optimal value, the number of jumps detected by the model is about $80\%$ of the real number of jumps. In contrast, values of $\Delta r$ and $\Delta t$ far from this optimal value result in a number of detected jumps far from the actual one, see Fig.~\ref{fig:synthetic}c. 
The results in Fig.~\ref{fig:synthetic} rely on the choice of $\delta r_\vibr$ (see \cite{SI}). For large $\delta r_\vibr$, vibrational noise dominates and the peak is suppressed (see \cite{SI}). In the opposite regime of small $\delta r_\vibr$, a spurious large peak of $\alpha$ appears at  small values of $\Delta t$. In this regime, the number of jumps is overestimated, since individual jumps can be counted multiple times for small $\Delta r$. We expect the occurrence of this regime to be due to a limitation of the synthetic model in accurately describing the dynamics at short (ballistic) time scales. Aside from this limitation, the synthetic model supports that $\alpha_{\max}$ well approximates the theoretical $\alpha$ associated with $P_\wait(t)$, and that the length and time scales $\Delta r_{\max}$, $\Delta t_{\max}$ associated with $\alpha_{\max}$ are comparable with the length and time scales of jumps.

\begin{figure}[htb]
\includegraphics[width=9cm]{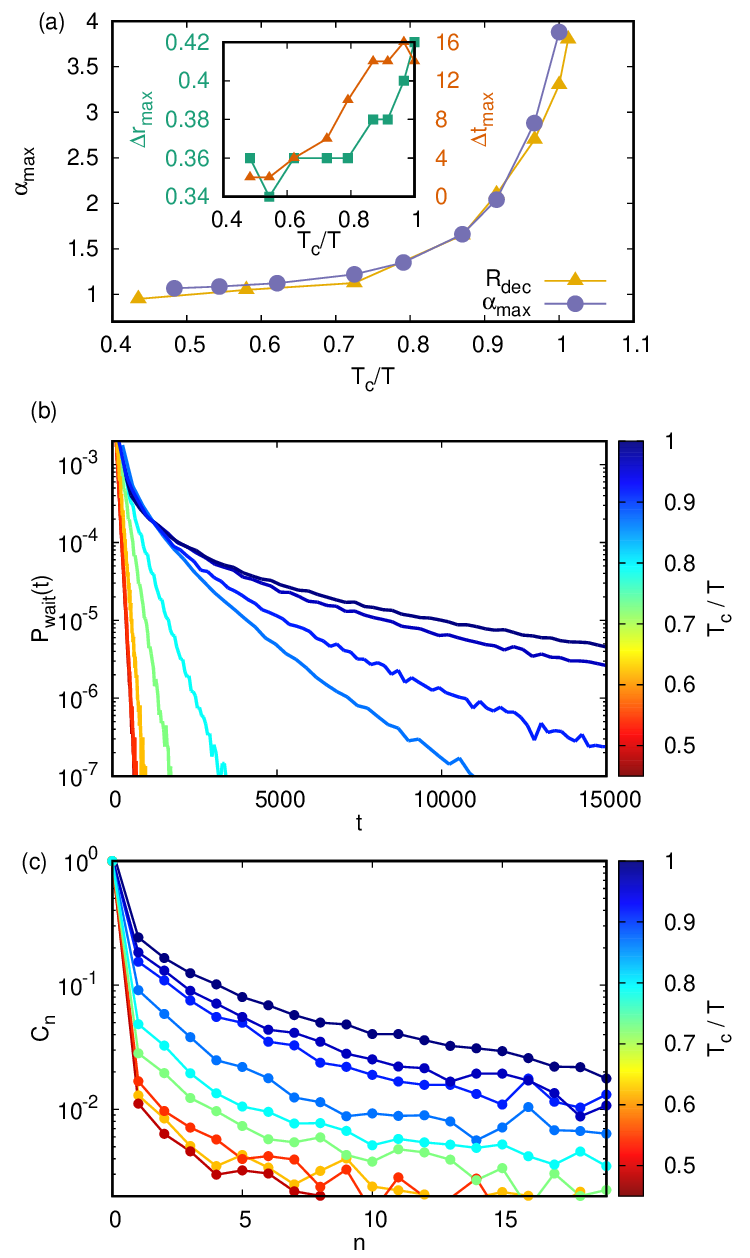}
\caption{(a) The values of $\alpha_{\max}$ (solid circles) match the translational decoupling $R_{\mathrm{dec}}$ extracted from Ref.~\cite{pinaki_non_gaussian} (solid triangles). Inset in a) $\Delta r_{\max}$ and $\Delta t_{\max}$ associated with $\alpha_{\max}$ as a function of temperature. Lines
joining the dots are to guide the eye. (b) Waiting time distributions associated with $\alpha_{\max}$. (c) Correlation functions  as defined in Eq.~\eqref{eq:correlation} for $\alpha_{\max}$.\label{fig:alpha-T}}
\end{figure}

We now analyze the behavior of $\alpha_{\max}$ as a function of temperature. At high temperature, $\alpha_{\max}$ is close to one, see Fig.~\ref{fig:alpha-T}a. This is consistent with the observation that $P_\wait(t)$ appears as nearly exponential in this regime, see Fig.~\ref{fig:alpha-T}b. Approaching $T_c$, $\alpha_{\max}$ substantially increases  (Fig.~\ref{fig:alpha-T}a) as the distribution of waiting times develops a longer tail (Fig.~\ref{fig:alpha-T}b). Both $\Delta r_{\max}$ and $\Delta t_{\max}$ tend to decrease at increasing temperature, see Fig.~\ref{fig:alpha-T}a, inset. This is consistent with the idea that, on increasing temperature, the dynamics become more homogeneous and $\alpha_{\max}$ is observed at shorter time windows, thus minimizing the number of undetected jumps. 

We find that the values of $\alpha_{\max}$ as a function of temperature closely match those of the translation decoupling $R_{\mathrm{dec}}$ reported in Ref.~\cite{pinaki_non_gaussian}, see Fig.~\ref{fig:alpha-T}a. The translational decoupling \cite{pre-garrahan,jcp-garrahan,jcp-garrahan-2,translational-decoupling-2,translational-decoupling-1} is an important observable, originally introduced in models describing the spatial dynamic heterogeneity of glassy systems \cite{Ediger-DH,Berthier-DH}. These models represent the system as a mosaic of regions constituted by spatially correlated sites  \cite{pre-garrahan,jcp-garrahan,translational-decoupling-2}. A particle moves when its site is reached by an excitation.  The initial particle persistence time corresponds to the structural relaxation time, i.e., the average time that a slow region should wait to be reached by an excitation. A second timescale is associated with self-diffusion, i.e., the average time between particle moves when they are in an excited, fast, region. As $T$ decreases, these two timescales decouple. The translational decoupling $R_{\mathrm{dec}}$ is the ratio of these two timescales. Without explicitly introducing spatial heterogeneity, Ref.~\cite{pinaki_non_gaussian} uses a CTRW model for the jump dynamics and shows that $R_{\mathrm{dec}}$, in the Kob-Andersen system, is comparable with the ratio between the time needed to observe the first jump and the time between all the subsequent jumps. The main difference between this result and that of Fig.~\ref{fig:alpha-T}a is that, in our case, $\alpha_{\max}$ is computed without making reference to CTRW, nor any other model, nor adjusting any free parameter. The close similarity of the two curves further corroborates that $\alpha_{\max}$ quantitatively captures an intrinsic property of the system, connected with its dynamical heterogeneity.

Since $\alpha$ only depends on the mean and variance of $P_\wait(t)$, correlations between waiting times in the time series do not affect our method. It is however interesting to see how these correlations change with temperature. For each temperature, we define \begin{equation}\label{eq:correlation}
 c_n=\sum_{i=1}^{N_A}\sum_{j=1}^{N_\wait-n}\frac{(t_\wait^{i,j}-\bar{t}_\wait)(t_\wait^{i,j+n}-\bar{t}_\wait)}{N_A(N_\wait-n)\sigma^2_{t_\wait}},
\end{equation}
where $n \in \mathbb{N}_0$, $t_\wait^{i,j}$ is the $j$-th waiting time in the trajectory of particle $i$ and $N_\wait$ the total number of waiting times along the particle trajectory. For all temperatures, $c_n$ shows an initial sharp drop followed by a slower decay  (Fig.~\ref{fig:alpha-T}c). At high temperature, correlations are almost suppressed, consistently with a homogeneous liquid dynamics. Approaching $T_c$, correlations become more pronounced.
Although our analysis does not require knowledge of these correlations, the slower decay of $P_\wait(t)$ and $c_n$ at low temperature are both manifestations of the increased dynamic heterogeneity, and can therefore be seen as two sides of the same coin.

We have presented a general procedure to identify jumps in glassy systems. This procedure neither relies on a model nor on adjustable parameters: the spatial and temporal scales defining a jump naturally emerge from the statistics of particle trajectories. Our method places the results of previous model-dependent approaches \cite{pinaki_non_gaussian,Ciamarra-jumps} in a broad, general framework. 

Our method is applicable to a broad class of systems from molecular and colloidal liquids to granular and biological materials \cite{berthier-biroli-review,Berthier_silica, pinaki_non_gaussian,Weeks_colloids,Walter_granular,MIT_lambda13,cell_migration-pnas,cytoplasm-cell}. Moreover,    current experimental techniques such as optical interferometry \cite{Optical-trapping-Inter-1,Optical-trapping-Inter-2} have reached nanometer spatial and microsecond temporal resolution and  revealed the ballistic motion of microsized particles \cite{Ballistic-Science}. Optical microscopy equipped with high speed cameras provides similar resolution \cite{SLM,Brightfield-Microscopy}. Thanks to these advancements, our method can be used to resolve hopping motion from experimentally observed trajectories.

\begin{acknowledgements}
We thank Juan P. Garrahan for feedback on a preliminary version of this manuscript.
\end{acknowledgements}

\bibliographystyle{apsrev}

\bibliography{Biblio}

\end{document}


\title{Supplementary Information: Inspection paradox and jump detection in glassy systems}

\author{Simone Pigolotti$^1$}
\email{simone.pigolotti@oist.jp}
\author{Sándalo Roldán-Vargas$^2$}%
\email{sandalo@ugr.es}
\affiliation{$^1$Biological Complexity Unit, Okinawa Institute of Science and Technology Graduate University, Onna, Okinawa 904-0495, Japan}
\affiliation{%
 $^2$Department of Applied Physics, Faculty of Sciences, University of Granada, 18071 Granada, Spain
}

\date{\today}

\maketitle

This document presents additional results that accompany the paper ``Inspection paradox and jump detection in glassy systems'' (from now on the ``Main Text''). We present in particular additional numerical results on the 1d model presented in Fig.~3 of the Main Text.\\

We here summarize the definition of the synthetic model, and present additional explanation on the choice of parameters. The dynamics is expressed by
\begin{equation}
\frac{dx}{dt}=v_\vibr+v_\jump\, ,
\end{equation}
where $v_\vibr$ and $v_\jump$ are velocities that describe the vibrational and hopping components of the motion, respectively.\\

The vibrational velocity is sampled from a double-exponential distribution: \begin{equation}
P(v_\vibr)=\frac{\delta t_\vibr}{2\delta r_\vibr}e^{-\frac{\delta t_\vibr}{\delta r_\vibr}|v_\vibr|}\, .
\end{equation}
The sampling of $v_\vibr$ is performed at regular interval of duration $\delta t_\vibr$. We fix $\delta t_\vibr=1$, on the order of the ballistic time of the Kob-Andersen system at the mode coupling temperature \cite{walter_andersen_PRE}. We shall see that the length scale $\delta r_\vibr$ is an important parameter. In Fig~3 of the Main Text, we have set $\delta r_\vibr=0.017$. With these choices, we have the effective diffusion coefficient $D_{\vibr}\equiv (\delta r_\vibr)^2/(2\delta t_\vibr)\approx 1.5\cdot 10^{-4}$. This value is comparable, but slightly higher than the diffusion coefficient of the Kob-Andersen system at the mode coupling temperature \cite{walter_andersen_PRE}.

\begin{figure}[htb]
\includegraphics[width=9cm]{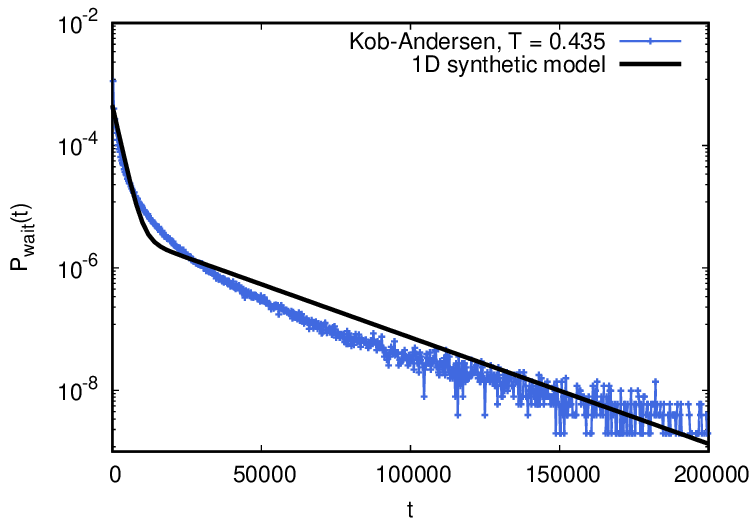}
\caption{Distribution $P_{\wait}(t)$ of intervals between jumps in the Kob-Andersen model at the critical temperature $T=0.435$ for $\alpha_{\max}$ (also shown in Fig.~4b in the Main Text), compared with the double-exponential distribution used in the 1d model $p_\jump(t)=p\exp(-t/t_1)+(1-p)\exp(-t/t_2)$ with $t_1=2000$, $t_2=25000$, and $p=0.9$.
\label{fig:synthetic_jumps}}
\end{figure}

The velocity $v_\jump$ is associated with hopping motion. The time intervals between jumps are sampled from double-exponential distribution $P_\wait(t)=p\exp(-t/t_1)+(1-p)\exp(-t/t_2)$. We set  $t_1=2000$, $t_2=25000$, and $p=0.9$ to mimic the distribution $P_\wait(t)$ that we observed in the Kob-Andersen system for $\Delta r_{\max}$ and $\Delta t_{\max}$ (see Fig.~\ref{fig:synthetic_jumps}). Each jump lasts for a duration $\delta t_\jump$; during these intervals, the hopping velocity is given by $|v_\jump|=\pm \delta r_\jump/\delta t_\jump$, where the sign is drawn with equal probability for each jump. We choose $\delta r_\jump=0.4$, which is on the order of $\Delta r_{max}$ (see Fig.2a in the Main Text). We choose $\delta t_\jump=2$, on the order of the time needed to travel ballistically a distance $\delta r_\jump=0.4$.

\begin{figure}[htb]
\includegraphics[width=15cm]{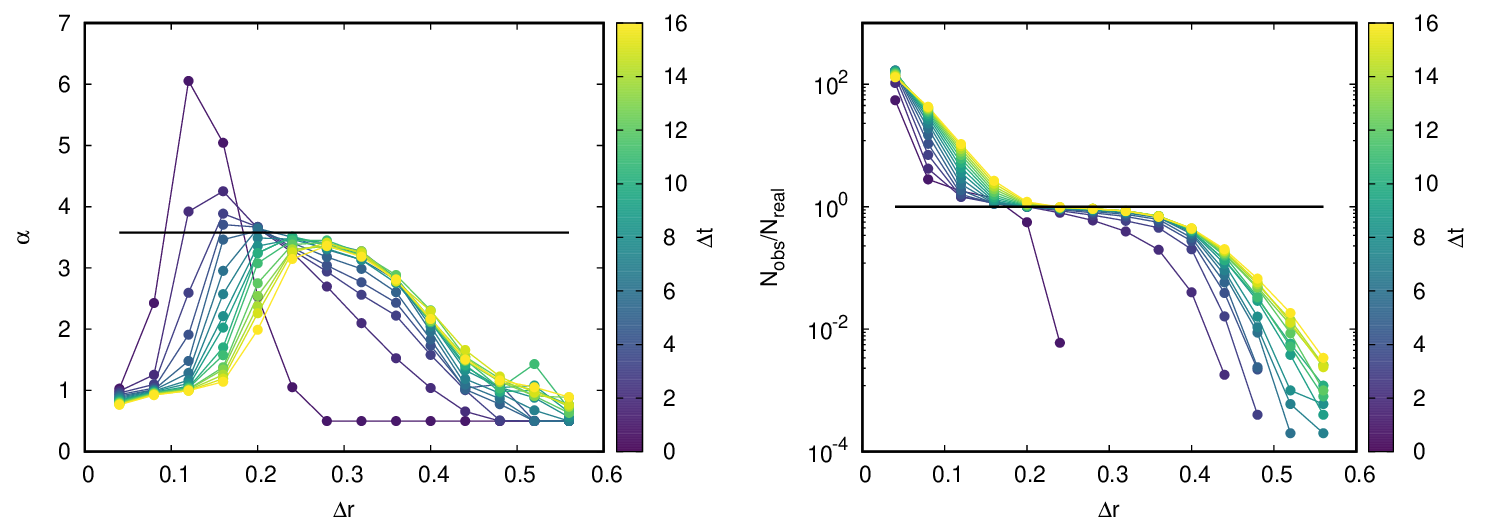}
\caption{Same as Fig.~3b and 3c in the Main Text, but with $\delta r_\vibr=0.01$
\label{fig:synthetic1}}
\end{figure}

\begin{figure}[htb]
\includegraphics[width=15cm]{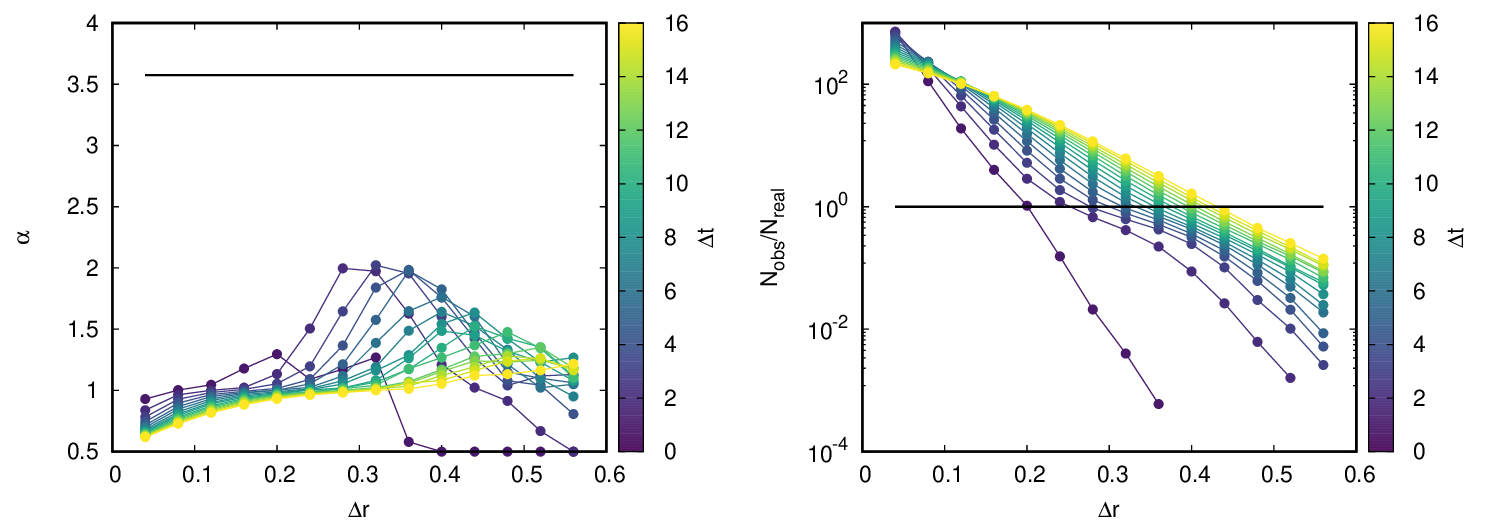}
\caption{Same as Fig.~3b and 3c in the Main Text, but with $\delta r_\vibr=0.024$
\label{fig:synthetic2}}
\end{figure}

The length scale $\delta r_\vibr$ is a crucial parameter for the model. For small  $\delta r_\vibr$, a spurious large peak in $\alpha$ appears for small $\Delta r$ and $\Delta t$, see Fig.~\ref{fig:synthetic1}. In this regime, each jump is detected multiple times, leading to a highly intermittent behavior and large $N_{\mathrm{obs}}/N_{\mathrm{real}}$, i.e., a large number of false positives. In the opposite regime of large  $\delta r_\vibr$, noise dominates. The peak in this case is at similar values of $\Delta r$ and $\Delta t$ as in the case of Fig.~3 of the Main Text. However, the value of $\alpha_{\max}$ is much lower than the theoretical value due to the disrupting effect of noise, see Fig.~\ref{fig:synthetic2}.

\bibliographystyle{apsrev}

\bibliography{si_Biblio}